\documentclass{aa}
\usepackage{graphics}
\usepackage{amssymb}
\begin{document}

\thesaurus{12(02.01.2;02.07.1;02.08.1)}
\title{A class of self-gravitating accretion disks}
\author{G. Bertin \and G. Lodato}
\offprints{G.Lodato}
\institute{Scuola Normale Superiore, Piazza dei Cavalieri 7, 56126 Pisa, Italy}
\date{received <date> / Accepted <date>}
\maketitle

\begin{abstract}
We consider a class of steady-state \\
self-gravitating accretion disks for
which efficient cooling mechanisms are assumed to operate so that the disk
is self-regulated at a condition of approximate marginal Jeans stability.
In an earlier paper, this scenario had been shown to lead naturally, in the
absence of a central point mass, to a self-similar solution characterized
by a flat rotation curve. In this article we investigate the entire
parameter space available for such self-regulated accretion disks and
provide two non-trivial extensions of the model. The first extension is
that of a bimodal disk, obtained by partially relaxing the self-regulation
constraint, so that full matching with an inner ``standard" Keplerian
accretion disk takes place. The second extension is the construction of
self-regulated accretion disks embedded in a diffuse spherical ``halo". The
analysis is further strengthened by a careful discussion of the vertical
structure of the disk, in such a way that the transition from self-gravity
dominated to non-gravitating disks is covered uniformly.

\end{abstract}

\section{Introduction}
For simplicity, traditional models of accretion disks ignore the
self-gravity associated with the accreting material and consider the case
where the disk is approximately Keplerian (e.g., see Pringle \cite{pringle}). 
On the
other hand, several studies have been carried out in order to investigate
the effects related to the disk self-gravity. Studies of this type have
addressed three different levels. The first stage is that where the disk
self-gravity is incorporated in the study of the vertical structure of the
disk. Here we may recall, among others, the pioneering work by
Paczy\'{n}ski (\cite{pacz}) and the very recent analysis by Bardou et al. 
(\cite{bardou}).
The second important level is that where one considers the effects of the
disk self-gravity on waves and on transport processes in the disk. Here we
may recall a number of interesting analyses, in particular the studies by Lin
\& Pringle (\cite{lin}), by Adams et al. (\cite{adams}), with the modified 
viscosity
prescription proposed by Lin \& Pringle (\cite{lin2}) (see also Andalib et al.
\cite{andalib} and references therein). Recent hydrodynamical
simulations (Laughlin \& Bodenheimer \cite{laugh}, Laughlin \& 
R\'{o}\.{z}yczka \cite{laugh2}, Laughlin et al. \cite{laugh3}) 
have focused on one key aspect of the modeling of accretion 
disks.
i.e. they have tried to assess whether the standard constant $\alpha$
viscosity prescription (Shakura \& Sunyaev \cite{shakura}) is justified in 
systems
where non-axisymmetric instabilities driven by the disk self-gravity play a
major role. The third level where the self-gravity of the disk can be
considered is that where, because of the gravitational field contributed by
the disk matter, the rotation itself of the accreting material is affected
and is no longer kept to be Keplerian. In this direction very few attempts
have been made (see, e.g., Bodo \& Curir \cite{bodo}).

The need for investigations where the self-gravity of the disk is properly
taken into account is further stimulated by recent accurate
observations that point to significant deviations from Keplerian rotation
in objects that are naturally interpreted as accretion disks (e.g., see
Greenhill et al. \cite{greenhill}, Moran \cite{moran}) and by the finding 
that in some contexts,
especially in the dynamics of protostellar disks, there are empirical
indications that the amount of mass in the disk is large (see  Hillenbrand et 
al. \cite{hillen}, Drimmel \cite{drimmel}). 
From the theoretical point of view, to address the dynamics of a 
self-gravitating
accretion disk is highly attractive especially because some {\it global} 
regulation aspects that
are emerging as interesting and relevant from various lines of thought (see
also Coppi \cite{coppi}) should find a natural manifestation in systems where 
the long-range forces are bound to generate an inherently global behavior.

The basic framework for the construction of models of accretion disks is
traditionally ``asymmetric". While the momentum transport equations are
generally replaced by a physically based prescription that bypasses our
ignorance on the detailed mechanisms that are involved (Shakura \& Sunyaev
\cite{shakura}; see also the recent modification proposed by Heyvaerts et al.
\cite{hey} and Bardou et al. \cite{bardou}), the energy
transport equations are usually kept in their ideal form, either following
a detailed inclusion of the radiation transport across the disk (e.g., see
Shakura \& Sunyaev \cite{shakura}; Bardou et al. \cite{bardou}) or by invoking
some ideal
``equation of state" and by arguing that the energy dissipated by viscosity
is partly redistributed in the disk (see Narayan \& Yi \cite{yi} and many
following articles). In a situation where we lack strong empirical
constraints (such as those that might be available in the laboratory) on
the detailed mechanisms at the basis of the various transport processes involved,
one might try to explore models where self-gravity, much like in galaxy
disks, plays a major role at all the three levels mentioned above, taking
the view that {\it both} momentum and energy transport equations should be
handled heuristically. This is indeed the step taken in an earlier
exploratory paper (Bertin \cite{bertin}), where the viscosity prescription 
suggested
by Shakura \& Sunyaev (\cite{shakura}) is retained, and, on a similar footing,
the energy transport equations are replaced by a physical condition of
self-regulation, related to marginal Jeans stability, as suggested by the
dynamics of galaxy disks (see Sect. 3).

In this paper we develop this latter point of view by describing the entire
parameter space available for {\it self-regulated accretion disks} (Sects. 2 
and 4), and by providing two non-trivial extensions of the model. The first
extension is that of a ``bimodal" disk, where full matching with an inner
``standard" Keplerian accretion disk is obtained (see Sect. 5.1). The
second extension is the construction of self-regulated accretion disks
embedded in a spherical diffuse ``halo", which may find application to the
extended disk associated with AGN's. The analysis is further strengthened
by a careful discussion of the vertical structure of the disk, in such a
way that the transition from self-gravity dominated to non-gravitating
disks is covered uniformly (Appendix A).
 
\section{Self-regulated accretion disks}
\label{section2}
\subsection{Basic equations and parameter space}

We consider a steady, axisymmetric, geometrically thin accretion disk, with
constant mass and angular momentum accretion rates, which we call 
$\dot{M}$
and $\dot{J}$, respectively. The conservation laws for mass and angular
momentum, in cylindrical coordinates, are:
\begin{equation}
\label{consmassa}
\dot{M}=-2\pi r \sigma u,
\end{equation}
\begin{equation}
\label{consangmoment}
\dot{J}=\dot{M}r^2\Omega+2\pi\nu\sigma r^3\frac{\mbox{d}\Omega}{\mbox{d}r},
\end{equation}
where $\sigma$ is the surface density of the disk, $u$ is the radial velocity,
$\Omega$ is the angular velocity, and $\nu$ is the viscosity coefficient.
Note that we take $\dot{M}>0$ in the case 
of inflow. Usually on the right hand side of
 Eq. (\ref{consangmoment}) the two terms have opposite signs; the first
term describes convection of angular momentum, while the second (negative in
the common situation where $d\Omega/dr<0$) accounts for the angular momentum 
transport due to viscous torques. Thus $\dot{J}>0$ corresponds to a net
angular momentum influx.

For a cool, slowly accreting disk, the radial balance of forces requires:
\begin{equation}
\label{eulero}
\Omega^2\sim\frac{1}{r}\frac{\mbox{d}\Phi_{\sigma}}{\mbox{d}r}+\frac{GM_{\star}}{r^3},
\end{equation}
where $M_{\star}$ is the mass of the central object and $\Phi_{\sigma}$ is
the disk contribution to the gravitational potential. The radial gravitational 
field generated by the disk can be written as:
\begin{equation}
					 \begin{array}{ll}
 &\displaystyle\frac{\partial\Phi_{\sigma}}{\partial r}(r,z)=\frac{G}{r}\int_0^{\infty}\left[K(k)-\frac{1}{4}\left(\frac{k^2}{1-k^2}\right)\times\right.\\
 &\displaystyle\left.\left(\frac{r'}{r}-\frac{r}{r'}+\frac{z^2}{rr'}\right) E(k)\right]\sqrt{\frac{r'}{r}}k\sigma(r')dr',
					\end{array}
\label{campovero}
\end{equation}
\noindent
where $E(k)$ and $K(k)$ are complete elliptic integrals of the first kind, and
\mbox{$k^2=4rr'/[(r+r')^2+z^2]$} (see Gradshteyn \& Ryzhik \cite{grad}). 
The field $\mbox{d}\Phi_{\sigma}/\mbox{d}r$ in the equatorial
plane is obtained by taking the limit $z\rightarrow 0$. Here we have preferred
to refer directly to the field (and not to the potential, as was done by 
Bertin \cite{bertin}) and to the formulae applicable to the case $z\neq 0$;
this representation is more convenient in view of the numerical investigations
that we have in mind (see Sects. 4 and 5).

For the viscosity we follow the standard prescription 
(Shakura \& Sunyaev \cite{shakura}):
\begin{equation}
\label{nu}
\nu=\alpha c h,
\end{equation}
where $c$ is the effective thermal velocity and $h$ the half-thickness of the
disk. Here
$\alpha$ is a dimensionless parameter, with $0<\alpha\lesssim1$; in the 
following we take it to be a constant. 

In the case of dominant disk self-gravity, one may adopt the requirement of 
hydrostatic equilibrium in the $z$ direction for a self-gravitating slab model,
which gives
\begin{equation}
\label{acca}
h=\frac{c^2}{\pi G\sigma}.
\end{equation}
A more refined analysis of the vertical equilibrium is provided in Appendix A.
Substituting Eq. (\ref{acca}) into Eq. (\ref{nu}) we obtain
$\nu\sigma=(\alpha/\pi G)c^3$,
which inserted in Eq. (\ref{consangmoment}) yields:
\begin{equation}
\label{angular}
G\dot{J}=G\dot{M}r^2\Omega+2\alpha c^3r^3\frac{\mbox{d}\Omega}{\mbox{d}r}.
\end{equation}

In order to close the set of equations (Eq. (\ref{eulero}), 
Eq. (\ref{campovero}), and Eq. (\ref{angular})), we need an additional 
relation. In standard studies this is
provided by an energy transport equation (see, for example, Pringle
\cite{pringle}; Narayan \& Popham \cite{popham} ; Narayan \& Yi \cite{yi}).
 Here we consider an alternative scenario (Bertin \cite{bertin}) where the
energy equation is {\it replaced} by
the requirement of marginal Jeans stability: 
\begin{equation}
\label{jeans}
\frac{c\kappa}{\pi G\sigma}=\bar{Q}\approx 1,
\end{equation}
where $\kappa$ is the epicyclic frequency. This assumes the presence of a
suitable {\it self-regulation mechanism} (see Sect. 3 below).
For a given value of $\alpha$ and $\bar{Q}$, we now have a complete set of
equations,
depending on the three parameters $M_{\star}$, $\dot{M}$, and $\dot{J}$.
We recall that the parameter $c\kappa/\pi G\sigma$ is the fluid analogue of the
axisymmetric stability parameter introduced and described by Toomre 
(\cite{toomre}) in his analysis of the stability of a stellar disk.

\subsection{The self-similar pure disk solution}
In the special case of $\dot{J}=0$ and $M_{\star}=0$, the problem is solved by
the self-similar solution with flat rotation curve (Bertin \cite{bertin}), 
characterized by: 
\begin{equation}
\label{mest}
2\pi G\sigma r=r^2\Omega^2=V^2=const.
\end{equation}
The other properties of the disk are specified by:
$c=(G\dot{M}/2\alpha)^{1/3}$,
$V=(2\sqrt{2}/\bar{Q})c$,
$u=-(\alpha \bar{Q}^2/4)c$,
$h/r=\bar{Q}^2/4$. Note that Eq. (\ref{mest}) describes the self-similar disks
found by Mestel (\cite{mestel}) in the non-accreting case.

The same solution is also valid asymptotically at large radii even when 
$M_{\star}$ and $\dot{J}$ do not vanish. In fact, the effect of a non-zero 
$M_{\star}$ should be unimportant for 
$$ r\gg r_s=2GM_{\star}(\bar{Q}/4)^2(G\dot{M}/2\alpha)^{-2/3}$$
 while the effects of a 
non-zero $\dot{J}$ should be unimportant for
$$r\gg r_J=\sqrt{2}(|\dot{J}|/\dot{M})(\bar{Q}/4)(G\dot{M}/2\alpha)^{-1/3}$$. 
Here the definitions of $r_s$ 
and $r_J$ are slightly different from those adopted earlier 
(Bertin \cite{bertin}).

\subsection{The iteration scheme for the general case}

For a given value of $\alpha$ and $\bar{Q}$ and for a specified accretion rate
$\dot{M}$, the general case is a problem with a well-defined physical 
lengthscale $r_s$ and one dimensionless parameter $\xi=sgn(\dot{J})r_J/r_s$.
In the following we wish to explore the properties of the mathematical 
solutions in the entire parameter space available (in particular, for both
positive and negative values of $\xi$). The knowledge of these solutions is a
prerequisite for a discussion of any specific astrophysical application. At a 
later stage, the physical problem under investigation is expected to restrict
the relevant parameter space. 

In order to calculate such one-parameter family of solutions, we start by 
writing the relevant equations in dimensionless form, in such a way that the
self-similar solution described above is easily recognized. We thus introduce
three {\it deviation functions} $\rho$, $\phi$, and $\chi$, which are related 
to the dimensional disk density $\sigma$, rotation curve $V$, and effective 
thermal speed $c$ by the following definitions:
\begin{equation}
\label{sigmaadimensionale}
2\pi G\sigma=\frac{V_0^2}{r}(1+\rho),
\end{equation}
\begin{equation}
\label{vadimensionale}
V^2=V_0^2\phi^2,
\end{equation}
\begin{equation}
\label{cadimensionale}
c=V_0\left(\frac{\bar{Q}}{2\sqrt{2}}\right)\chi,
\end{equation}
with $V_0^2=(8/\bar{Q}^2)(G\dot{M}/2\alpha)^{2/3}$. Clearly the functions
$(1+\rho)$, $\phi^2$, and $\chi$ are all positive definite; the self-similar
solution corresponds to $\rho=0$, $\phi=1$, $\chi=1$. 

In terms of the dimensionless radial coordinate $x=r/r_s$, the basic set of 
equations (corresponding to Eqs.(\ref{eulero}), (\ref{campovero}),
(\ref{angular}), and (\ref{jeans})) thus become:
\begin{equation}
\label{velocita}
\phi^2=1+\hat\phi^2[\rho/x]+\frac{1}{x}
\end{equation}
\begin{equation}
\label{velocita1}
				\begin{array}{ll}
 &\displaystyle\hat\phi^2[\rho/x]=\frac{1}{2\pi}\int_0^{\infty}dx'\frac{k}{\sqrt{xx'}}\times\\
 &\displaystyle\left[K(k)-\frac{1}{4}\frac{k^2}{1-k^2}\left(\frac{x'}{x}-\frac{x}{x'}+\frac{\delta^2}{xx'}\right)E(k)\right]\rho(x'),
				\end{array}
\end{equation}
\begin{equation}
\label{chi}
\chi(x)=\left(1-\frac{\xi}{x\phi}\right)^{1/3}\left(1-\frac{\mbox{d}\ln\phi}{\mbox{d}\ln x}\right)^{-1/3},
\end{equation}
\begin{equation}
\label{sigma}
\rho(x)=\chi\phi\sqrt{1+\frac{\mbox{d}\ln\phi}{\mbox{d}\ln x}}-1,
\end{equation}
with $k^2=4xx'/[(x+x')^2+\delta^2]$, in the limit $\delta\rightarrow 0$.

In the case of outward angular momentum flux ($\xi<0$), the above equations can
be readily solved by iteration in the following way. An initial seed solution
$\phi_0=\sqrt{1+1/x}$ is inserted in Eqs. (\ref{chi}) and (\ref{sigma}), thus
leading to a first approximation to the density deviation $\rho_0(x)$. This is
inserted in Eq. (\ref{velocita1}), then producing (via Eq. (\ref{velocita})) a
new expression for the rotation curve deviation $\phi_1(x)$. Typically three or
four iterations are sufficient to reach a satisfactory convergence.

For the case when the net angular momentum flux is inwards ($\xi>0$), 
Eq. (\ref{chi}) shows that we may run into a difficulty at small radii, where
the quantity $(1-\xi/x\phi)$ changes sign.
Here we may proceed by analogy with standard studies (e.g., see Pringle 
\cite{pringle}), by restricting our analysis to the outer disk defined by 
$x>x_{in}$, with the condition $x_{in}\phi(x_{in})=\xi$, i.e. 
$\dot{J}=\dot{M}r_{in}^2\Omega(r_{in})$. This is taken to occur at a point
where the angular velocity $\Omega(r)$ reaches a maximum (so that
$\mbox{d}\ln\phi/\mbox{d}\ln x=1$ and $\chi$ may remain finite); such maximum 
is identified as a location, in the vicinity of the surface of the central 
object, to which the accretion disk is imagined to be connected by a relatively
narrow boundary layer. Of course, in the boundary layer the physical processes 
will be different from those described in our model (for example, there will be
effects due to pressure gradients and the disk will be thick). Thus we will not
be able to follow the associated (inwards) decline of the angular velocity away
from the 
Keplerian profile. As a result, in our calculation we let the effective thermal
speed vanish at $r_{in}$;
this unphysical behavior would be removed when one describes the boundary
layer with more realistic physical conditions (see Popham \& Narayan 
\cite{pophamboundaries}).
\label{chiboundary}

\section{Self-regulation in self-gravitating accretion disks}

Before proceeding further, we should make here a digression and try to
better explain the physical justification at the basis of the regulation
prescription set by Eq. (\ref{jeans}), which is the distinctive property of the
class of models addressed in this paper. This discussion expands and clarifies 
a brief description provided earlier (Bertin \cite{bertin}).

\subsection{The mechanism}

We start by noting that, when self-gravity dominates, processes associated
with the Jeans instability basically determine the relevant scales of
inhomogeneity of the system under consideration. On the other hand, it is
well known (Toomre \cite{toomre}) that in the plane of a thin, 
self-gravitating,
rotating disk the tendency toward collapse driven by self-gravity, which is
naturally expected to be balanced locally only at small wavelengths by
pressure forces, can be fully healed by rotation also at small wavenumbers.
As a result, if the disk is sufficiently warm, it can be locally stable
against all axisymmetric disturbances. The condition of marginal stability
thus ensured is usually cast in the form $Q = 1$, where $Q$ is proportional
to the effective thermal speed of the disk.

Therefore, in galactic dynamics it has long been recognized that an
initially cold disk is subject to rapid evolution, on the dynamical
timescale. For a stellar disk, the Jeans instability induces fast overall
heating of the disk up to levels of the effective thermal speed for which
the instability is removed, basically following the above criterion ($Q
\approx 1$). The heating rate is thus very sensitive to the instantaneous
value of $Q$. Therefore, even in the absence of dissipation, for relatively
low values of $Q$ the collective Jeans instability provides a mechanism to
stir the system and to heat it. However, a collisionless non-dissipative
disk would be unable to evolve in the opposite direction, i.e. to cool if
initially hot ($Q > 1$), and, because of other residual processes (such as
tidal interactions), may still be subject to some perennial heating even
when Jeans instability is ineffective. In turn, if efficient dissipation is
available (such as that associated with the interstellar medium in galaxy
disks), a competing mechanism can cool an initially hot disk down toward
conditions of marginal Jeans instability.

These important dynamical ingredients have been studied from various points
of view, giving rise to interesting scenarios, also by means of numerical
experiments (see Miller et al. \cite{miller}, Quirk \cite{quirk}, Quirk \& 
Tinsley \cite{tinsley}, Sellwoood \& Carlberg \cite{sellwood}, Ostriker 
\cite{ostriker}), so that eventually the
possibility of an actual {\it self-regulation} in self-gravitating disks
has been formulated and explored. Accordingly, the two competitive
mechanisms, of dynamical heating and dissipative cooling, can set up a kind
of dynamical thermostat. In particular, it has been shown that
self-regulation is very important in determining the conditions for the
establishment of spiral structure in galaxies (Bertin and Lin \cite{bertinlin};
and references therein). The general concept has also found successful
application to the dynamics of the interstellar medium, even when
dynamically decoupled from the stellar component, in relation to the
interpretation of the observed star formation rates (Quirk \cite{quirk2}; 
Kennicutt \cite{kennicutt}). Note that the most delicate aspect of the 
regulation process is the
cooling mechanism; in galaxy disks, an important contribution to cooling is
thus provided by cloud-cloud inelastic collisions that dissipate energy on
a fast timescale.

A final remark is in order. As shown by the case of galaxy disks, the dynamics
involved may be extremely
complex, so that there is little hope to provide a simple ``ideal" energy
equation for the description of a coupled disk of stars (of different
populations) and gas (in different forms and phases). This point has an 
additional important consequence. If we try to describe such an inherently
complex system by means of an idealized one-component model, we should be 
ready to introduce the use of some \emph{effective} quantities (in particular, 
of an effective thermal speed; see also the use of the term ``effective'' after
Eq. (\ref{nu}) and at the beginning of this subsection).

\subsection{Viability of the mechanism for some accretion disks}

It should be emphasized that the main strength of the self-regulation
scenario described above is rooted in semi-empirical arguments. The actual
data from many galaxy disks (among which the Milky Way Galaxy) and from
planetary rings show that conditions of marginal stability ($Q \approx 1$)
often occur and can indeed be established in systems subject to complex
dynamical processes. For accretion disks, there may already be some
empirical indications in this direction, to the extent that the application
of standard models to some observed systems points to very cold disks, with
$Q$ well below unity (e.g., see Kumar \cite{kumar}, as briefly mentioned in
Sect. 6). Additional clues in the same direction also derive from numerical
experiments; in the context of protostellar disks, numerical simulations
show that disks formed under conditions where self-gravity is important
include wide regions characterized by a constant $Q$-profile (Pickett et al.
\cite{pickett}).

An accretion disk may be subject to efficient cooling by a variety of
mechanisms, depending on the physical conditions that characterize the
specific astrophysical system under consideration. From this point of view,
the cooling necessary for the establishment of self-regulation may occur
efficiently already via the radiative processes included in ``standard"
models (see the general discussion of the outer disk by Bardou et al.
\cite{bardou}, which we will summarize in Sect. 5.1). In reality, the
dynamics of matter slowly accreting in a disk can be significantly more
complex. Cold systems, such as a protogalactic disk, a protostellar disk,
or the outer parts of the disk in an AGN, may have a composite and complex
structure. They may include dust, gas clouds, and other particulate objects
with a whole variety of sizes and ``temperatures". Much like for the HI
component of the interstellar medium, the main contribution to the
effective temperature of the disk might be from the turbulent speed of an
otherwise cold medium. On the one hand, for these systems it may be hard or
even impossible to write out a simple ``ideal" energy transport equation.
On the other hand, such a complex environment is likely to possess all the
desired cooling and heating mechanisms that cooperate in self-regulation.
In this respect, one is thus encouraged to bypass the problem of defining a
representative set of equations for energy transport, and to use instead
the semiempirical prescription of Eq. (\ref{jeans}). Somewhat in a similar way,
our
inability to derive from first principles a satisfactory set of equations
for momentum transport is often taken to justify the adoption of the
$\alpha$-prescription of Eq. (\ref{nu}). These phenomenological 
prescriptions have
several limitations, but may still work as a useful guide to our efforts
and provide interesting models to be compared with the observations.

To be sure, some types of accretion disk, or some regions inside accretion
disks (for example, very close to the center; see Sect. 5.1), may lack the
physical ingredients invoked above. In fact, there is no reason to claim
that self-gravity must {\it always} be important. Therefore, we will study
the structure of self-regulated accretion disks, as a viable class of
astrophysical systems, while we do recognize that warmer, non-regulated
disks may exist and are likely to be basically free from the effects
associated with the self-gravity of the disk.

\subsection{The impact of $Q$ on momentum and energy transport}

The self-regulation mechanism has been demonstrated by considering a
simplified set of equations (Bertin \cite{bertiniau}) where efficient cooling 
is
included and the role of self-gravity is modeled by means of a heating term
with an analytic expression (inversely proportional to a high power of $Q$)
meant to incorporate the results of dynamical studies that show that
heating is indeed very sensitive to the value of Q. The main features of this
formula, with its threshold at $Q \approx 1$, aimed at representing the
``thermal evolution" of the disk, are somewhat analogous to the heuristic
characterization of the viscosity dependence on $Q$ adopted by Lin \&
Pringle (\cite{lin2}) in the parallel problem of constructing the momentum
transport equations when self-gravity is important.

In our discussion of self-regulated accretion disks, we actually have no
doubt that self-gravity is likely to have an important impact on viscosity,
and this is still tacitly incorporated in the $\alpha$ prescription. This
impact is even more obvious if one recalls that a self-gravitating disk can
be subject to non-axisymmetric instabilities, which are bound to contribute
significantly to angular momentum transport. Our class of axisymmetric,
steady-state accretion models represents only one approximate idealization
of the actual system that we are addressing. Given the indications of
several dynamical studies (in addition to those of Lin \& Pringle, see, for
example, Laughlin \& Bodenheimer \cite{laugh}, Laughlin \& R\'{o}\.{z}yczka
\cite{laugh2}), a more complete analysis should thus include one further 
relation
between $\alpha$ and $Q$. In general, this might practically require that
the phenomenological prescription (\ref{nu}) be used with a parameter $\alpha$
varying with radius. In reality, we believe that the proper way to include
the relevant physical effects, especially those associated with
non-axisymmetric instabilities, would be through some {\it global}
constraint. Until such global description remains not available, the
assumption of a free, constant $\alpha$ may provide a first approximation,
best applicable when $Q$ is self-regulated. This choice can be physically
consistent {\it a posteriori}, at least for the self-similar solution of
Sect. 2.2.

\section{Properties of models with a central point mass}
\label{section3}
After this digression, we can now proceed with the analysis of the problem as
formulated in Sect. 2.
In the presence of a central point mass ($M_{\star}\neq 0$) we expect the
lengthscale $r_s$ to mark the transition from a Keplerian disk to a fully
self-gravitating disk with flat rotation curve. Surprisingly, in our class of
self-regulated accretion disks the role of the disk self-gravity turns out to 
be significant all the way down to the center.

\subsection{Rotation curves}
\begin{figure}
  \resizebox{\hsize}{!}{\includegraphics{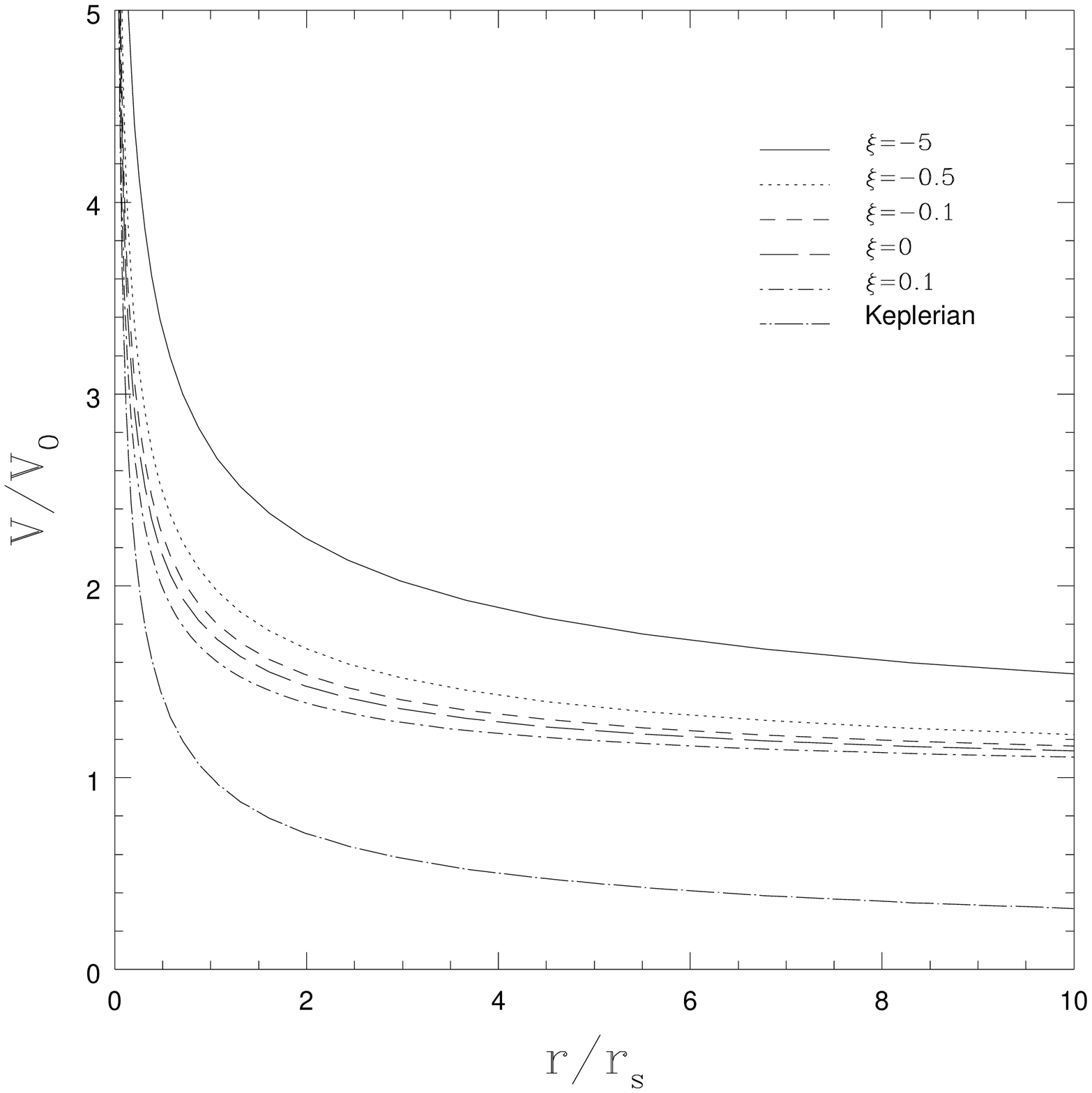}}
  \resizebox{\hsize}{!}{\includegraphics{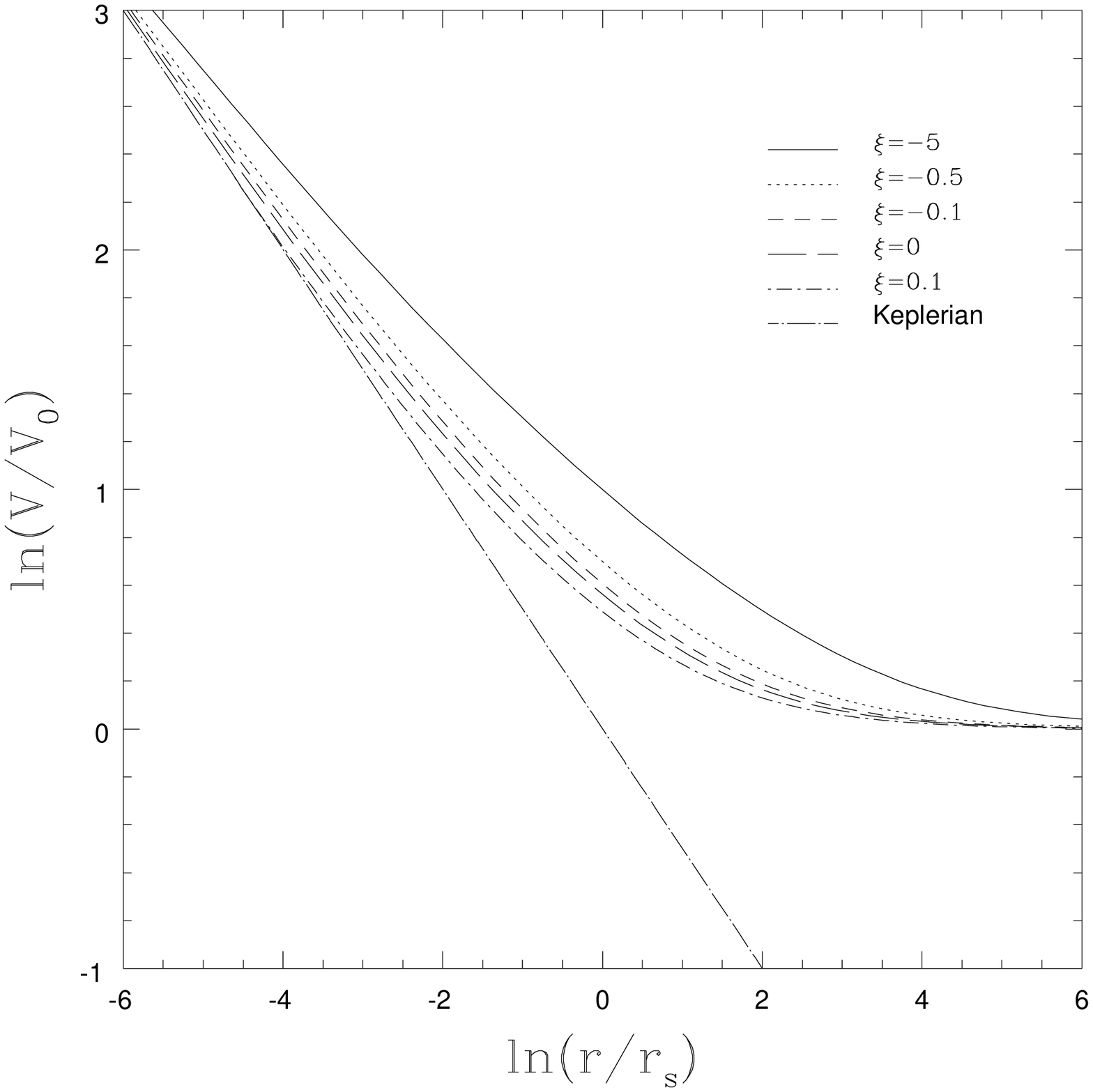}}	
  \caption{\small{{\bf Top panel:} Rotation curve of the disk in the presence 
of a central point mass for different values of the angular momentum flux 
parameter. {\bf Bottom panel:} Same as top panel, but in logarithmic scale.}}
  \label{fig:rotazione}
\end{figure}

In Fig.~\ref{fig:rotazione} we illustrate the behavior of the rotation 
curves in our class of models for several values of the angular momentum flux
parameter $\xi$. For comparison, we also show the Keplerian curve $V_K$ that is
obtained by setting $1+\rho=0$. We note that the difference from the Keplerian
decline is significant even at $r=O(r_s)$. For example, for the $\xi=0$ case we
find $(V-V_K)/V_K\approx 100\%$ at $r=2r_s$.

\subsection{Effective thermal speed}
\begin{figure}
  \resizebox{\hsize}{!}{\includegraphics{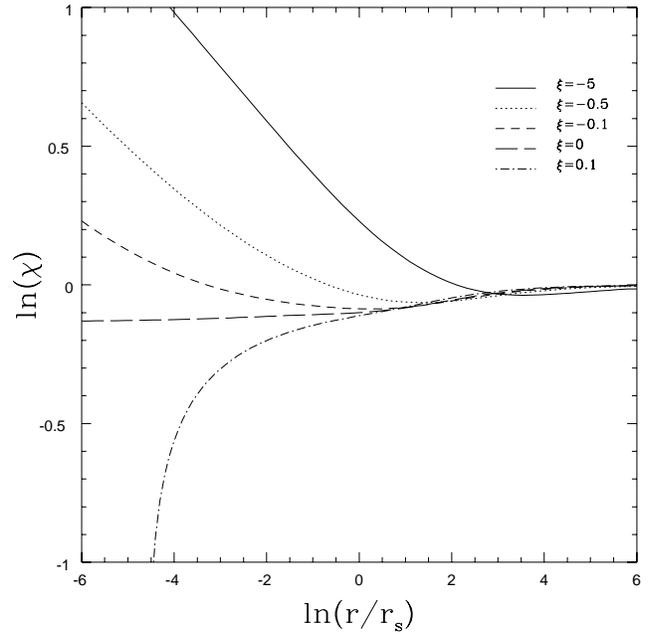}} 
  \caption{\small{Equivalent thermal speed of the disk. The two cases ($\xi>0$ 
and $\xi<0$) show opposite behavior in the inner disk.}}
  \label{fig:termica}
\end{figure}  

Disks where the angular momentum is transported outwards ($\xi<0$) tend to 
develop a warmer core, while the opposite trend occurs for disks where the
angular momentum is carried inwards ($\xi>0$). This is shown in 
Fig.~\ref{fig:termica} (Fig.~\ref{fig:termica} and the following 
Fig.~\ref{fig:massa}, Fig.~\ref{fig:epsilon}, and Fig.~\ref{fig:spessore} are 
shown in logarithmic scale to better bring out the 
behavior in the inner parts of the disk). Note that in the case of inward 
angular momentum flux
($\xi>0$) the effective thermal speed need not vanish at the inner
edge of the disk (as it does in our models) where 
a boundary layer is expected to be generated (see discussion at the end of 
Section \ref{chiboundary}).

\subsection{Role of the disk self-gravity close to the central point mass}
\begin{figure}
  \resizebox{\hsize}{!}{\includegraphics{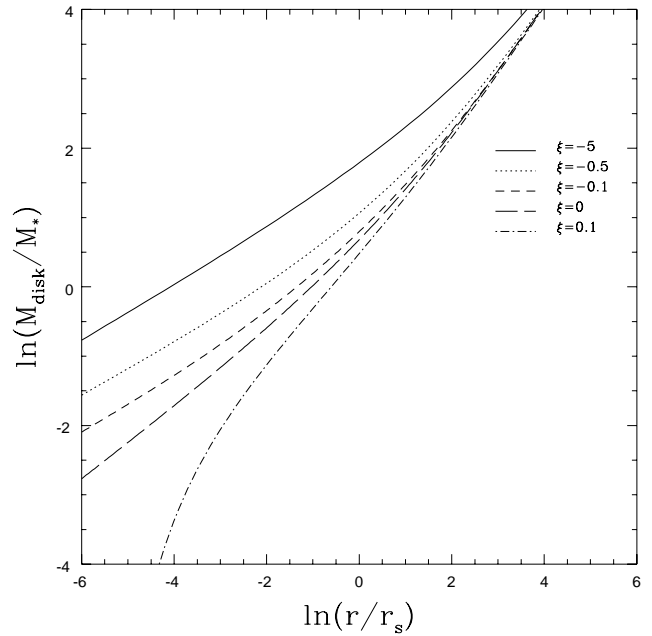}} 
  \caption{\small{Cumulative mass of the disk relative to that of the central
object.}}
  \label{fig:massa}
\end{figure}
\begin{figure}
  \resizebox{\hsize}{!}{\includegraphics{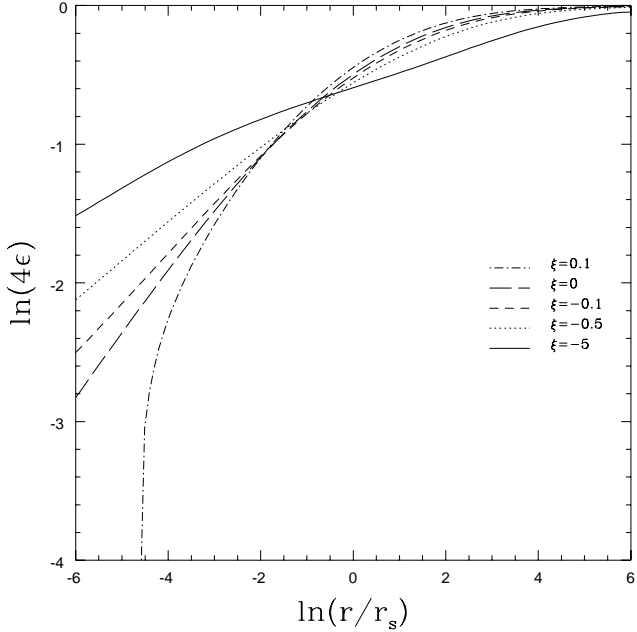}} 
  \caption{\small{Local self-gravity of the disk, as measured by 
$\epsilon=\pi G\sigma/r\kappa^2$, for different values of the angular momentum
flux parameter.}}
  \label{fig:epsilon}
\end{figure}
There are several quantities directly related to the disk density distribution
in the disk that allow us to characterize the role of the disk self-gravity.

The most natural quantity to consider is the ratio of the mass of the disk to
that of the central object. Obviously, for our non-truncated models this
quantity is meaningful only when referred to a given radius. Fig.~\ref{fig:massa}
shows how rapidly in radius the system becomes dominated by
the mass of the disk. Note that, in any case, 
$M_{disk}(r)/M_{\star}\rightarrow 0$ for $r\rightarrow 0$.

In galactic dynamics the local disk self-gravity is usually measured in terms
of the parameter $\epsilon=\pi G\sigma/r\kappa^2$. The fully self-gravitating
self-similar disk (with flat rotation curve) is characterized by 
$\epsilon=1/4$. The profile of this parameter (see Fig.~\ref{fig:epsilon})
confirms that indeed, close to the center, the influence of the central mass
becomes stronger and stronger.

\begin{figure}
  \resizebox{\hsize}{!}{\includegraphics{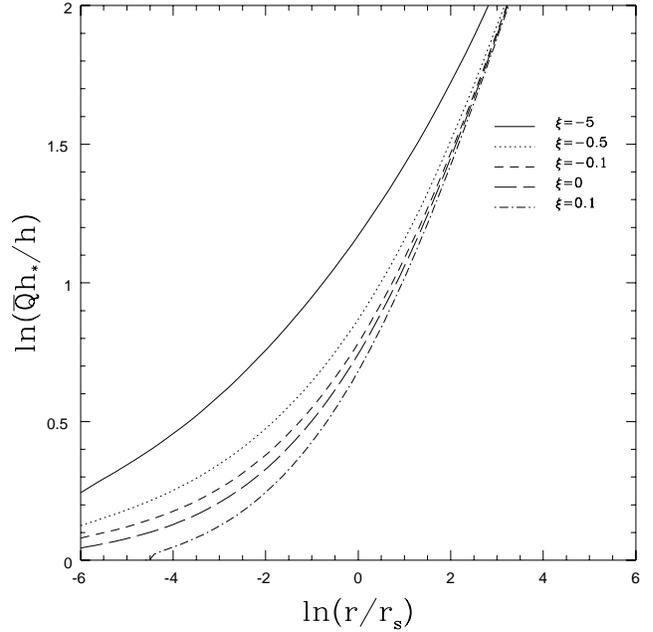}} 
  \caption{\small{Ratio of the vertical scaleheight $h_{\star}=cr/V_K$ to the
thickness of the disk for various models.}}
  \label{fig:spessore}
\end{figure}

Given the behavior of the profiles $M_{disk}(r)/M_{\star}$ and $\epsilon(r)$,
one might conclude that the innermost disk should be treated as a standard 
Keplerian accretion disk. This conclusion is contradicted by the following
argument. In setting up the equations of our models, we have taken the 
vertical equilibrium to be dominated by the disk self-gravity (see 
Eq. (\ref{acca})). If the innermost disk were fully Keplerian, at small radii
the vertical scaleheight $h_{\star}=cr/V_{K}$ should become much smaller than
the thickness $h$ associated with our models. Instead, a plot of the ratio
$h_{\star}/h$ only shows that the two scales become {\it comparable} to each
 other
(see Fig.~\ref{fig:spessore}), thus demonstrating that the influence of the
disk self-gravity is significant all the way to the center. This, of course,
reflects our choice of imposing the self-regulation prescription 
(Eq. (\ref{jeans})) at all radii; in the next section we will show that the
disk can indeed make a true transition to a Keplerian disk if the 
self-regulation prescription is suitably relaxed.

In view of the above considerations, in order to check that no major 
consequences arise from the use of a vertical equilibrium prescription only
partly justified for $r<r_s$, we have also considered models based on the
 improved 
prescription for the disk thickness (see Appendix A):
\renewcommand{\arraystretch}{2.8}
\begin{equation}
\label{improve}
					\begin{array}{ll}
h= & \displaystyle \frac{c^2}{\pi G\sigma}
      \frac{\pi}{4Q^2(2\Omega^2/\kappa^2-1)}\times\\
   & \displaystyle \left[\sqrt{1+\frac{8}{\pi}Q^2\left( \frac{2\Omega^2}{\kappa^2}-1\right)}-1\right]. 
					\end{array}
\end{equation}
Note that in the limit $2\Omega^2/\kappa^2\rightarrow 1$, applicable to the
self-similar disk, this prescription reduces to Eq. (\ref{acca}). In the 
innermost region $\Omega\rightarrow V_K/r$, $\Omega/\kappa\rightarrow 1$, so 
that $h_{\star}/h\rightarrow (4\bar{Q}/\pi)/[\sqrt{1+8\bar{Q}^2/\pi}-1]$.

\begin{figure}
  \resizebox{\hsize}{!}{\includegraphics{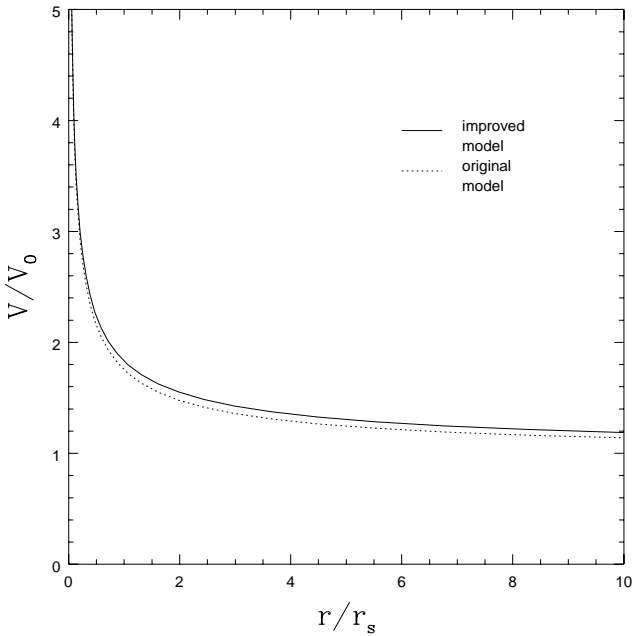}} 
  \caption{\small{Rotation curves for two different treatments of the vertical
structure for the case $\xi=0$, $\bar{Q}=1$.The solid line represents the 
improved model, based on 
Eq. (\ref{improve}); the dotted line is the original model,
based on Eq. (\ref{acca}).}}
  \label{rcorr}
\end{figure}

In Fig. \ref{rcorr} we show the rotation curve of the improved model, compared
to that of the original one, for the $\xi=0$ case. This plot (along with 
similar results for the other relevant physical quantities of the disk)
shows that, qualitatively,
the more refined vertical analysis leaves the models basically unchanged.

\section{Extensions}
\subsection{Matching with an inner Keplerian, non-self-regulated accretion 
disk}

If we take an astrophysical situation (such as an AGN or a protostar) with
specific physical conditions, it is likely that the arguments that support
the adoption of the self-regulation constraint, followed in this paper,
fail outside a well-defined radial range, either at small or at large
radii. For example, we may recall that in the context of the dynamics of
spiral galaxies the relevant $Q$ profile is argued to be flat in the outer
disk, but is thought to increase inwards inside a circle of radius $r_Q$
often identified as the radial scale of influence of the bulge; even in the
absence of a bulge, the central parts of the disk are thought to be
generally hotter (e.g., see Bertin \& Lin, \cite{bertinlin}). In our context, we may
then imagine an accretion disk where at small radii self-regulation fails,
$Q$ becomes large, and, correspondingly, the disk self-gravity ceases to be
important.

This discussion suggests that it should be interesting to explore the
possibility where Eq. (\ref{jeans}) is replaced by a condition of the form
$c\kappa/\pi G \sigma = Q(r)$, with a $Q$ profile that decreases
monotonically with radius and reaches the self-regulation value $\bar{Q}$
only beyond a certain radial scale $r_Q$. What would be the impact of such
a choice on the structure of the accretion disk? What might be the physical
arguments leading to the justification of such a profile? In particular,
what would set the scale $r_Q$ for given values of the parameters $\alpha$,
$M_{\star}$, $\dot{M}$, and $\dot{J}$? Note again that we are reversing the
standard point of view, whereby one may ask what is the $Q$ profile for an
accretion disk, based on the structure calculated from a given choice of
the energy equations.

Imposing, as we are going to do, a given profile $Q(r)$ may, at first sight,
appear to be arbitrary. In reality, the freedom in the choice of the profile
allows us to test quantitatively how the dynamical characteristics $V(r)$,
$M_{disk}(r)$, $\epsilon(r)$, and $h(r)$ of the disk change when the 
self-regulation constraint is partially relaxed, in the inner disk, in a
variety of ways. It is up to us to test the different possibilities (which may
correspond to completely different sets of energy balance equations in the 
inner disk) that might be considered. For our purposes, since we wish to
study the deviations from the standard ``Keplerian'' case, we only need to take
into account that the relevant physical processes match so that in the outer
disk self-regulation is enforced. Note that in the transition region where
matching between the inner and the outer disk occurs, for reasons expressed in
Sect. 3, it may be practically impossible to define, from first principles, a
satisfactory set of energy equations able to include all the desired radiation
processes and the non-linear effects associated with Jeans instability. 

This procedure draws considerable support from a recent analysis of
``standard" disks (Bardou et al. \cite{bardou}) aimed at detecting evidence 
for the importance of disk self-gravity in the outer disk. Based on an
extension of the ``standard" $\alpha$-disks (characterized by Kramers'
opacity and by neglect of radiation pressure), the
effects of the disk self-gravity have been here incorporated by means of an 
improved thickness prescription
(somewhat in the spirit of our Appendix A) and of a modified
viscosity prescription, but the (Keplerian) $\Omega$ profile is left
unaltered. Therefore, this study is ideally suited to describe the
conditions of our inner disk, as we intend to partially relax the
self-regulation requirement. A very important result of the analysis by
Bardou et al. (\cite{bardou}) is that their ``standard" description breaks down
beyond a radius $r_Q$, well inside which the local stability parameter behaves
approximately as $Q \sim r^{-9/8}$; as it might have been anticipated, the
location where the standard model breaks down coincides with the location
where $Q$ becomes approximately equal to unity. In conclusion, the analysis
by Bardou et. al (\cite{bardou}) encourages us to consider the following choice
of $Q$ profile

\begin{equation}
\label{qprofile}
\frac{c \kappa}{\pi G \sigma} = \bar{Q} \{1 + (r/r_Q)^{-9/8} \exp[-(r/r_Q)]\}
\end{equation}

\noindent to be used instead of Eq. (\ref{jeans}). (The formula is meant to be
used as a semi-empirical tool; one should keep in mind that the exact form of
the $Q$ profile will be determined by the detailed energy processes
occurring in the inner disk and by the progressively important role of 
collective instabilities.) If we express the scale $r_Q$ found in that
study in terms of our scale $r_s$, we find

\begin{equation}
\frac{r_Q}{r_s}\simeq 0.8\cdot10^{-4}\frac{\alpha^{-2/45}}{\bar{Q}^2}\left(\frac{M_{\star}}{10^8M_{\odot}}\right)^{-2/3}\left(\frac{\dot{M}}{10M_{\odot}/yr}\right)^{8/45}.
\end{equation}

\noindent Note that the ratio $r_Q/r_s$ {\it decreases} while $M_{\star}$
increases, and that its value is only weakly dependent on $\alpha$ and on
$\dot{M}$. For parameters typical of an AGN, self-regulation may thus be
ensured very far in, while an extrapolation of the above recipe to
parameters typical of a disk surrounding a T Tauri star suggests that, for
these latter objects, $r_Q$ should become $O(r_s)$.

\begin{figure}
  \resizebox{\hsize}{!}{\includegraphics{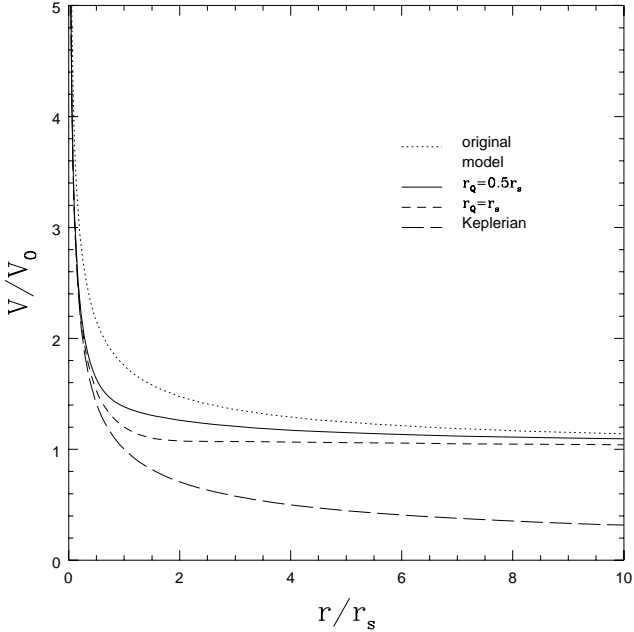}} 
  \caption{\small{Rotation curves of partially self-regulated models with the 
$Q$-profile given in Eq. (\ref{qprofile}), for different values of $r_Q$. The
thickness prescription used is that of Eq.~(\ref{improve}).}}
  \label{fig:qprofile}
\end{figure}

Some examples of {\it partially self-regulated models} computed on the
basis of Eq. (\ref{qprofile}) (see Fig. \ref{fig:qprofile}) show how an outer
disk dominated by
self-gravity can match in detail with an inner standard Keplerian accretion
disk.

\subsection{The effect of a diffuse ``halo''}
So far we have considered the case where the mass is all distributed in a
disk (either at the center, as a point mass, or in diffuse form). In view
of possible applications to AGN configurations or to the general galactic
context, it is important to consider an extension of the models to the case
where part of the gravitational field is determined by a diffuse spherical
component, which we will call {\it halo} (even if it may just correspond to
the central region of an elliptical galaxy). This will lead to rotation
curves otherwise not accessible by our models. On the other hand, it is
easily recognized that this natural extension is going to leave the slow
density decline of the disk unaltered. Therefore, if we are interested in
producing models with finite total mass, we should be ready to impose an
outer truncation radius or, which might effectively be equivalent, to relax
the self-regulation prescription in the outermost disk.

\begin{figure}
  \resizebox{\hsize}{!}{\includegraphics{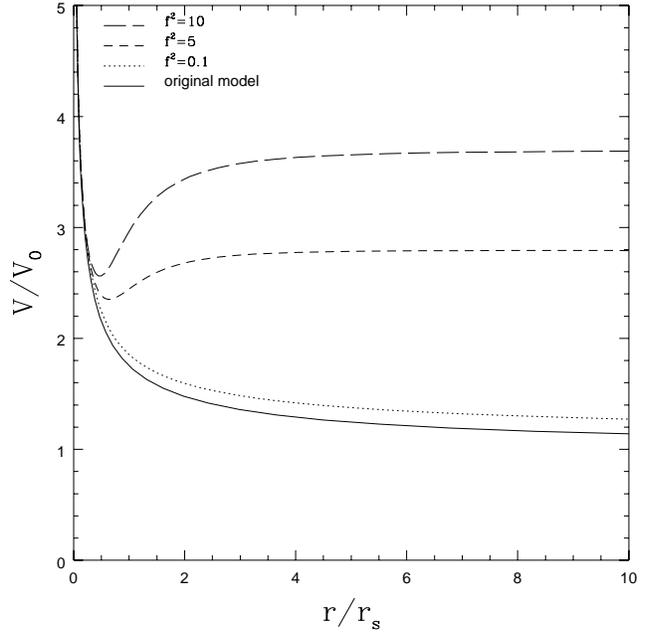}} 
  \caption{\small{Rotation curves of models with a diffuse halo, with $\xi=0$,
and core radius equal to $r_s$ ($x_0=1$), for different values of $f^2$.}}
  \label{fig:alone}
\end{figure}
We have thus considered a set of models where the field external to the disk
is produced by the joint contribution of a central point mass and of a halo 
(which, for simplicity, we take to
be spherical). In view of the case of a disk embedded in an
elliptical galaxy, we have modeled the halo as approximately isothermal, with
a finite core radius. In this case the dimensionless 
equation giving the rotation curve (Eq. (\ref{velocita})) is modified as 
follows:
\begin{equation}
\phi^2=1+\frac{1}{x}+\hat\phi^2[\rho/x]+\frac{f^2x^2}{x_0^2+x^2}.
\end{equation}
We see that
now the equations depend on two additional parameters: $f^2$, giving 
the relative strength of the external field, and $x_0$, which measures the size
of the core radius. 
In this case it is easy to demonstrate that at large radii the density 
deviation $\rho$ approaches $f^2$ if $f\ll 1$, and $f$ if $f\gg 1$.

In Fig. \ref{fig:alone} we show examples of the rotation curve of 
models with a diffuse halo, for the case $\xi=0$, $x_0=1$. For the vertical 
structure, we have referred to the improved vertical prescription of 
Eq.~(\ref{improvedprescription}), with $\Omega^2_{ext}\propto f^2/(x_0^2+x^2)$.

\subsection{Disks with an outer truncation radius}

Self-regulated accretion disks with finite mass can be easily constructed
by imposing the existence of an outer truncation radius. Either the study
of the collapse of a gas cloud with finite mass or the consideration of the
physical conditions in the outer parts of some astrophysical objects will
naturally bring us to address such models.

\section{Discussion}

In this paper we have developed a framework for the construction of a class
of self-gravitating accretion disks, independently of the specific
conditions characterizing the astrophysical systems where the accretion
disk paradigm applies. In principle, we might even speculate that the above
framework could be the starting point to describe some stages of the
formation of protogalactic disks (e.g. see Fall \& Efstathiou \cite{fall}) or
the establishment of some regular extended HI disks in elliptical galaxies
(see Morganti et al. \cite{morganti}). In practice, our class of models appears
to be best applicable to two categories of astrophysical objects, i.e. to
Active Galactic Nuclei and to protostellar disks. Concrete applications,
which will require a detailed consideration of the available observational
constraints, are beyond the scope of the present paper. Still we would like
to make a few comments that should illustrate why the models appear to be
promising for the above categories of objects.

Recently, from radio maser emission, it has been possible to obtain
accurate measurements of the rotation of the disk in the central parts of a
few Active Galactic Nuclei; these include \object{NGC 4258}, \object{NGC 1068},
\object{Circinus}, \object{NGC 4945}, \object{NGC 3079}, and \object{NGC 1386}
(see Moran, \cite{moran}). 
These measurements have shown that, although in some cases (as for 
\object{NGC 4258}, see
Miyoshi et al. \cite{miyoshi}) the rotation curve is Keplerian to a high degree
of approximation, there may be significant deviations from the $r^{-1/2}$
profile. For example, in \object{NGC 1068} the rotation velocity at
distances $\approx 1$ pc from the ``center" turns out to decline as
$r^{-0.35}$ (see Greenhill et al. \cite{greenhill}; Kumar \cite{kumar}). The 
general
conclusion is that rather extended disks, which may contain large amounts
of mass, are present; indeed one possible reason for the deviation from the
Keplerian decline has been identified in the influence of the gravitational
field contributed by the accretion disk itself. At the same time, the
application of ``standard" accretion disk models to some of these objects
has led to finding values of the $Q$ parameter well below unity (for 
\object{NGC 1068}, see Kumar \cite{kumar}). All these are clues that show that 
a model where the
role of the disk self-gravity is fully incorporated is called for. Note
that if we adopt the numbers suggested by the data for \object{NGC 1068} we
would find $r_s \approx 25$ pc and $r_Q \approx 1.6~10^{-4} r_s$; then, it is
curious to find that a self-regulated disk with these characteristics would
have, for $\xi = - 5$, a gradient of the rotation curve at $r \approx 1$ pc
compatible with $V \sim r^{-0.37}$.

On a completely different mass scale, if we refer to the case of
protostellar disks, typically quoted parameters are $\dot{M} \approx
10^{-8} M_{\odot}/yr$ and $M_{\star} \approx 0.5 M_{\odot}$ (Hartmann et
al. \cite{hartmann}). Under these circumstances we find $r_s \approx 1000$ AU 
and $r_Q \approx 0.8 r_s$. Interestingly, protostellar disks have been 
observed to
extend out to a radius from $\approx 100$ AU to $\approx 1000$ AU (Dutrey
et al. \cite{dutrey}; McCaughrean \& O'Dell \cite{dell}). As for the case of 
AGN's, a
check on the values of the temperatures anticipated on the basis of the
effective thermal speeds predicted by the self-regulated models shows that
the numbers fall reasonably within the range suggested by the observations.

Finally, it is interesting to note that studies of the \emph{spherical, 
inviscid} collapse of a molecular cloud, imagined to eventually generate a 
system composed of a protostar and a circumstellar disk, lead to mass accretion
rates $\dot{M}\propto c^3$ (Hunter \cite{hunter}, Shu \cite{shu}), curiously 
analogous to those of our 
completely different accretion scenario. This coincidence may offer interesting
clues for modeling the entire process where a cloud, starting out in an 
initially spherical collapse, ends up in a protostellar accretion disk.
\begin{acknowledgement}
We would like to thank Nicola Attico for interesting discussions. This work is
partially supported by MURST and by ASI of Italy.
\end{acknowledgement}
\appendix
\section{Modeling the vertical density profile}

The vertical structure of the disk is generally derived by imposing hydrostatic
equilibrium in the $z$-direction under the assumption of isothermality.
The relevant Poisson equation and hydrostatic equilibrium condition for an 
axisymmetric configuration can be written as (here the symbol $\rho$  denotes 
volume mass density and
should not be confused with the density deviation function of the main text):
\begin{equation}
\label{poisson}
\frac{1}{r}\frac{\partial}{\partial r}\left(r\frac{\partial \Phi}{\partial r}\right)+\frac{\partial^2\Phi}{\partial z^2}=4\pi G\rho+4\pi G\rho_{ext},
\end{equation}
\begin{equation}
\label{hydro}
\frac{c^2}{\rho}\frac{\partial\rho}{\partial z}=-\frac{\partial\Phi}{\partial z},
\end{equation}
where we have taken polar cylindrical coordinates.
For simplicity, we assume that the external field associated with $\rho_{ext}$
is spherical. [In the class of models studied in Sect.~\ref{section2} and 
Sect.~\ref{section3} we have taken the case where
 $\rho_{ext}=M_{\star}\delta({\bf x})$.]
From the definitions of $\Omega$ and of $\kappa$, it is readily shown that,
close to the equatorial plane, Eq.~(\ref{poisson}) can be rewritten as:

\begin{eqnarray}
\label{poisson2}
\displaystyle\frac{\partial^2\Phi}{\partial z^2} &=&4\pi G\rho+4\pi G\rho_{ext}+2\Omega^2-\kappa^2\\
			    &=&4\pi G\rho+(2\Omega^2-\kappa^2)_{\sigma}+\Omega_{ext}^2.
\label{bahcallext} 	
\end{eqnarray}

The latter expression follows from the exact relation $4\pi G\rho_{ext}=(\kappa^2-\Omega^2)_{ext}$
applicable to a spherical density distribution. The expression $(2\Omega^2-\kappa^2)_{\sigma}$
indicates that the quantity in parentheses should be calculated based on the
radial field given by Eq.~(\ref{campovero}). For a relatively thin disk the
quantity $(2\Omega^2-\kappa^2)_{\sigma}+\Omega^2_{ext}$ may be taken to be
nearly independent of $z$, so that the gravitational field obtained by 
integrating Eq.~(\ref{bahcallext}) includes a term that is approximately linear
in $z$. The result can be inserted in the right hand side of Eq. (\ref{hydro}),
which becomes:
\begin{equation}
\label{eq:bahcall}
\frac{\mbox{d}^2\sigma_z}{\mbox{d}z^2}=-\frac{\mbox{d}\sigma_z}{\mbox{d}z}\left\{\frac{2\pi G}{c^2}\sigma_z+\left[\frac{(2\Omega^2-\kappa^2)_{\sigma}}{c^2}+\frac{\Omega_{ext}^2}{c^2}\right]z\right\},
\end{equation}
where we have introduced the integrated density $\sigma_z=2\int_0^z\rho(z')\mbox{d}z'$.
Note that the term $\Omega_{ext}^2z$ is just the vertical component of the 
spherical external field, as might have been anticipated. If $\rho_0=\rho(z=0)$
is the density on the equatorial plane, we can define two scales,
$\Sigma=\sqrt{2\rho_0/\pi G}c$ and $H=c/\sqrt{2\pi G\rho_0}$,  and the natural 
dimensionless parameter $A=[(2\Omega^2-\kappa^2)_{\sigma}+\Omega^2_{ext}]/4\pi G\rho_0$,
so that Eq. (\ref{eq:bahcall}) becomes:
\begin{equation}
\label{eq:spessadim}
\frac{\mbox{d}^2y}{\mbox{d}\zeta^2}=-2\frac{\mbox{d}y}{\mbox{d}\zeta}\left[y+A\zeta\right],
\end{equation}
to be solved under the conditions $y(0)=0$, $y'(0)=1$; here we have used the 
dimensionless variables $y=\sigma_z/\Sigma$ and $\zeta=z/H$.
For each value of $A$ one can compute the desired density profile and the
associated surface density $\sigma=\sigma_z(z=\infty)$. One can then introduce
a scaleheight $h$ such that $\sigma=2\rho_0h$, and the value of $h$ can be
computed directly from the relation $h=Hy(z=\infty)$.

The standard non self-gravitating model, where the contribution of the gas 
density $\rho$ to the vertical gravitational field is negligible, corresponds
to the limit equation $y''=-2A\zeta y'$. Usually the analysis is carried out
with the additional approximation $A\approx \Omega_{ext}^2/4\pi G\rho_0$, but
this is not needed; by retaining the contribution $(2\Omega^2-\kappa^2)_{\sigma}$
one can thus keep track of the effects associated with the rest of the disk 
mass distribution, which may be significant even where the local disk density
$\rho_0$ is small. In any case, such limit equation yields the well known
Gaussian profile $\rho=\rho_0\exp(-z^2/2h_{\star}^2)$, with 
$h_{\star}^2=H^2/2A=c^2/[(2\Omega^2-\kappa^2)_{\sigma}+\Omega^2_{ext}]$, here
improved with respect to the standard expression $h_{\star}\approx c/\Omega_K$
used in the so-called Keplerian limit. Note that $h_{\star}$ is slightly 
different from the scaleheight $h$ defined above, since $h=\sqrt{\pi/2}h_{\star}$.

The limit of the homogeneous fully self-gravitating slab corresponds to the
equation $y''=-2yy'$, so that the vertical density profile is given by
(Spitzer \cite{spitzer}) $\rho=\rho_0~\mbox{sech}^2(z/h)$, with
$h=c^2/\pi G\sigma$; here the scale $h$ is the same as that defined by the 
relation $\sigma=2\rho_0h$.
Note that even when no external (spherical) field is present, the solution for 
an axisymmetric disk obtained from Eq.(\ref{eq:spessadim}) is not exactly
the one derived in the homogeneous, self-gravitating slab, unless the 
rotation curve is flat.
\begin{figure}
  \resizebox{\hsize}{!}{\includegraphics{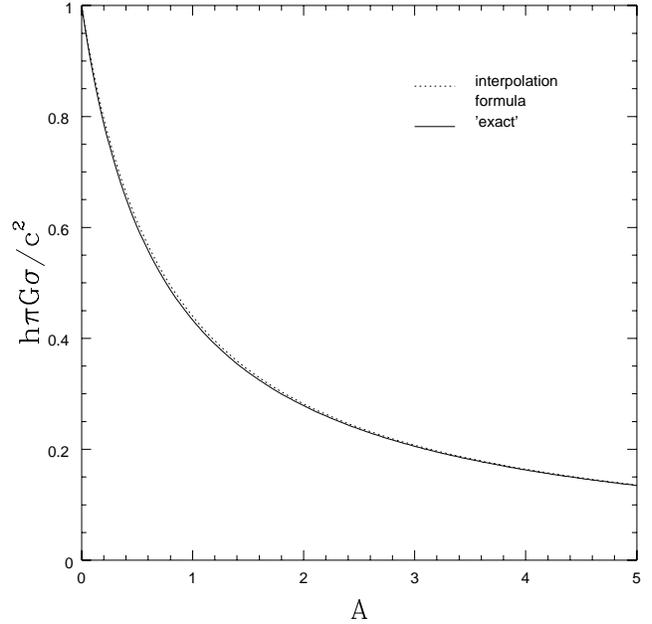}} 
  \caption{\small{Comparison between the exact (numerically computed based on 
Eq.~(\ref{eq:spessadim})) and approximate (Eq.~\ref{eq:interpol}) expression 
for the disk thickness.}}
  \label{fig:interpol}
\end{figure}

The density profile associated with Eq.(\ref{eq:spessadim}) is neither
Gaussian nor $\mbox{sech}^2$. For practical purposes it may be convenient to 
use a simple interpolation formula for the vertical scale, which is justified
by the following description ``biased'' towards the Gaussian limit. If we start
from Eq.(\ref{hydro}), naively expand the vertical field as 
$(\partial\Phi/\partial z)(z)\sim (\partial^2\Phi/\partial z^2)_{z=0}z$, and
then use Eq.(\ref{poisson}) to estimate the factor $(\partial^2\Phi/\partial z^2)_{z=0}$,
we find an equation leading to an unrealistic Gaussian density profile with 
scaleheight $h$ given by:
\begin{equation}
\frac{\pi}{2h^2}=\frac{4\pi G\rho_0}{c^2}[1+A].
\end{equation}

The quantity $\rho_0$ contains a dependence on $h$ (for given $\sigma$).
This suggests the use of the following interpolation formula:
\begin{equation}
\label{eq:interpol}
h=\left(\frac{c^2}{\pi G\sigma}\right)\frac{1}{1+4A/\pi}.
\end{equation}

Note that the exact calculation, from $h=Hy(z=\infty)$ gives a relation
$h=(c^2/\pi G\sigma)f(A)$, with $f(A)=[y(z=\infty)]^2$. The choice of the
factor $(4/\pi)$ in Eq.~(\ref{eq:interpol}) guarantees the proper limits for
$A=0$ and for $A\rightarrow\infty$. The accuracy of the interpolation formula
is illustrated in Fig.~(\ref{fig:interpol}). In turn, since $A$ depends on $h$
and on $\sigma$ via the quantity $\rho_0$, for the purposes of the present 
paper it may be convenient to reexpress Eq.~(\ref{eq:interpol}) as:
\begin{equation}
\label{improvedprescription}
h=\frac{c^2}{\pi G\sigma}\frac{\pi}{4a}\left(\sqrt{1+8a/\pi}-1\right),
\end{equation}
with $a=Q^2[(2\Omega^2-\kappa^2)_{\sigma}+\Omega^2_{ext}]/\kappa^2$. This leads
to Eq.~(\ref{improve}) of the main text, applicable when the external field
is produced by a central point mass.
This interpolation formula improves on earlier analyses (Sakimoto \& Coroniti
\cite{sakimoto}, Bardou et al. \cite{bardou}), in several respects, allowing us 
to better describe the transition between a Keplerian and a self-gravity
dominated disk and to extend the treatment to the case where the external 
field is not just that of a simple point mass at the center.
\label{app}

\end{document}